\documentclass[letterpaper]{article} % DO NOT CHANGE THIS
\usepackage{aaai2026}  % DO NOT CHANGE THIS  
\usepackage{times}  % DO NOT CHANGE THIS
\usepackage{helvet}  % DO NOT CHANGE THIS
\usepackage{courier}  % DO NOT CHANGE THIS
\usepackage[hyphens]{url}  % DO NOT CHANGE THIS
\usepackage{graphicx} % DO NOT CHANGE THIS
\usepackage{natbib}  % DO NOT CHANGE THIS AND DO NOT ADD ANY OPTIONS TO IT
\usepackage{caption} % DO NOT CHANGE THIS AND DO NOT ADD ANY OPTIONS TO IT
\usepackage{amssymb}
\usepackage{amsmath}
\usepackage{amsthm}
\usepackage{booktabs}

\usepackage{algorithm}
\usepackage{algorithmic}
\newtheorem{theorem}{Theorem}
\newtheorem{definition}{Definition}
\usepackage{times}
\usepackage{helvet}
\usepackage{courier}
\usepackage{xcolor}

\usepackage{newfloat}
\usepackage{listings}
\DeclareCaptionStyle{ruled}{labelfont=normalfont,labelsep=colon,strut=off} % DO NOT CHANGE THIS
\floatstyle{ruled}
\newfloat{listing}{tb}{lst}{}
\floatname{listing}{Listing}
\title{Untraceable DeepFakes via Traceable Fingerprint Elimination}
\author{
    Jiewei Lai\textsuperscript{\rm 1},
    Lan Zhang\textsuperscript{\rm 1},
    Chen Tang\textsuperscript{\rm 1},
    Pengcheng Sun\textsuperscript{\rm 1},
    Xinming Wang\textsuperscript{\rm 1},
    Yunhao Wang\textsuperscript{\rm 2},
}
\affiliations{
    \textsuperscript{\rm 1}University of Science and Technology of China, Hefei, China\\
    \textsuperscript{\rm 2}Lenovo Research, Beijing, China\\
    jw\_lai@mail.ustc.edu.cn,zhanglan@ustc.edu.cn,\{chentang1999,speical0806,xinmingwang\}@mail.ustc.edu.cn,wangyh43@lenovo.com
}

\begin{document}

\maketitle

\begin{abstract}
Recent advancements in DeepFakes attribution technologies have significantly enhanced forensic capabilities, enabling the extraction of traces left by generative models (GMs) in images, making DeepFakes traceable back to their source GMs.
Meanwhile, several attacks have attempted to evade attribution models (AMs) for exploring their limitations, calling for more robust AMs.
However, existing attacks fail to eliminate GMs' traces, thus can be mitigated by defensive measures.
In this paper, we identify that untraceable DeepFakes can be achieved through a multiplicative attack, which can fundamentally eliminate GMs' traces, thereby evading AMs even enhanced with defensive measures.
We design a universal and black-box attack method that trains an adversarial model solely using real data, applicable for various GMs and agnostic to AMs.
Experimental results demonstrate the outstanding attack capability and universal applicability of our method, achieving an average attack success rate (ASR) of 97.08\% against 6 advanced AMs on DeepFakes generated by 9 GMs.
Even in the presence of defensive mechanisms, our method maintains an ASR exceeding 72.39\%.
Our work underscores the potential challenges posed by multiplicative attacks and highlights the need for more robust AMs.
\end{abstract}

    % \link{Code}{https://anonymous.4open.science/r/TEST-F4B1}
    % \link{Datasets}{https://aaai.org/example/datasets}
    % \link{Extended version}{https://aaai.org/example/extended-version}

\section{Introduction}
With the rapid development of GMs, such as generative adversarial networks (GANs)~\cite{DBLP:conf/iclr/KarrasALL18,DBLP:conf/iclr/MiyatoKKY18,DBLP:conf/iclr/BinkowskiSAG18} and diffusion models (DMs)~\cite{DBLP:conf/cvpr/RombachBLEO22}, creating realistic and diverse high-quality images is becoming easily accessible.
However, this progress also enables misuse, leading to issues such as misinformation and intellectual property infringement. Therefore, dedicated research efforts are being paid to forensics, such as DeepFakes detection for authenticity identification ~\cite{DBLP:journals/csur/WangLCLW25}.
% ~\cite{DBLP:conf/ijcai/LiuZS023,DBLP:conf/ijcai/Bao0WLWJC24,DBLP:conf/ijcai/Shi00MHW24}

DeepFakes attribution, a method that surpasses DeepFakes detection by identifying both the authenticity of an image and the specific model or type of models used to generate it, is a promising approach for enhancing accountability among malicious content creators.
This technology captures distinctive traces left by GMs within images, known as the model fingerprint, thereby attributing generative content to its source model.
Additionally, it supports intellectual property protection by identifying unauthorized use of copyrighted models or their generative content.

% 直接说脆弱性研究的重要性（cite）+现有工作怎么做+现有工作不足
With significant advances in DeepFakes attribution, research into the vulnerability of AMs, known as attribution attacks, has emerged to explore their limitations, thereby guiding the development of more robust attribution methods and effective countermeasures. 
Current techniques evade AMs by adding perturbations into images, called additive attacks, demonstrating considerable attack performance.
However, our analysis and preliminary experiments show they are easy to defend against because they fail to alter or eliminate the traceable fingerprints that are essential for attribution.
This implies that while existing methods can temporarily circumvent attribution, their inability to eliminate traceable fingerprints renders them inherently fragile and circumventable due to persistent forensic traces.

Therefore, this paper further explores attack methods against attribution methods, developing an attack strategy to achieve untraceable DeepFakes via fingerprint elimination that not only evades AMs but also circumvents defense mechanisms.
To achieve this objective, three critical issues must be addressed. 
% 第一点说明修改是针对模型指纹进行修改
1) The trade-off between fingerprint elimination and visual imperceptibility. While more extensive modifications generally enhance attack performance, they can also compromise image quality. Moreover, failing to eliminate the fingerprint undermines the effectiveness of the attack. 
2) The broad spectrum of GMs and their diverse fingerprint characteristics present significant complexity. It is impractical to design a model-specific method, therefore, ensuring the universality of our approach enhances its applicability to DeepFakes generated by various GMs without requiring customization. 
3) Additionally, the adversary often lacks knowledge of the attribution mechanisms in practical scenarios.
% 改成模型无关
Thus, designing a model-agonist attack that is independent of specific AMs is essential for ensuring its effectiveness in evading various unknown AMs.

To address these challenges, we design a universal and black-box attack strategy targeting AMs and defensive mechanisms.
In contrast to an additive attack, grounded in the principle of DeepFakes attribution, we theoretically identify that the multiplicative attack fundamentally evades attribution by 
% 不要让攻击者觉得直接把乘性攻击拿过来就能，强调我们做了什么/贡献
eliminating fingerprints within DeepFakes through an adversarial matrix. Specifically, we establish the theoretical existence of such matrices and a lower bound on multiplicative attack success probability.
% 直接说防御困难+原因
Critically, our analysis demonstrates that defending against multiplicative attacks is fundamentally challenging in practice: inverting adversarial images to recover original content requires access to adversarial images and corresponding original DeepFakes simultaneously, and incurs prohibitive computational complexity, making reliable defense practically unattainable.
% rendering multiplicative attacks provably robust against AMs and defensive mechanisms.
Subsequently, we propose a universal and black-box multiplicative attack framework to achieve untraceable DeepFakes using exclusively real data without requiring any DeepFakes or GMs.
Specifically, this framework comprises two modules:
1) Data synthesis: this module employs sampling and transformation units to create synthetic data with real data, mimicking the characteristics of DeepFakes.
2) Model Construction: within this module, a comprehensive loss function is designed for jointly optimizing in the perceptual, spatial, and frequency domains, guiding the trained model to eliminate artificial traces within synthetic images during training.
3) Fingerprint Elimination:
% 直观的介绍攻击流程
Give DeepFakes generated by any GMs, the resulting model effectively serves as the adversarial matrix to eliminate fingerprints within DeepFakes, thus evading unknown AMs even with the presence of the defensive mechanisms.

% We conduct an in-depth evaluation of our proposed method using six advanced DeepFakes attribution technologies on DeepFakes generated by six GANs and two DMs. 
% The results demonstrate the effectiveness and universality of our approach.
% Our method achieves the highest attack success rate (ASR) across all AMs, with an average ASR of 99.81\%, surpassing state-of-the-art (SOTA) attack methods.
% Moreover, black-box and white-box defensive measures are applied, and the results demonstrate that our attack method achieves significant effectiveness with an ASR of more than 87.89\% against these defensive mechanisms.
% Quantitative analysis in both frequency and image domains confirms that our method can effectively eliminate traceable fingerprints within DeepFakes with multiplicative operations.

Our main contributions are summarized as follows:
% 前两点在文章篇幅少，考虑合成一点？
\begin{itemize}
\item  We demonstrate that current attacks against AMs are fundamentally constrained to additive perturbations through analysis and preliminary experiment, this additive nature inherently preserves model-specific fingerprints and renders them highly susceptible to effective defenses.
\item  We identify that the multiplicative attack can provably eliminate model fingerprints and establish their theoretical existence and a lower bound on success. We further show they are irreversible, thereby proving the intrinsic evasiveness against AMs, even under defenses.
\item We propose a universal and black-box multiplicative attack framework to achieve untraceable DeepFakes using only real data. By synthesizing fingerprint-mimicking data and optimizing a multi-domain loss, the model learns to effectively eliminate the fingerprints, enabling evasion of unknown AMs across diverse GMs.
\item We experimentally validate our method's effectiveness across 6 advanced AMs and 9 GMs, achieving an ASR of 97.08\%, surpassing SOTA methods.  Additionally, our method demonstrates significant effectiveness with an ASR of more than 72.39\% against both black-box and white-box defensive mechanisms.  
\end{itemize}

\section{Related Works}

\subsubsection{DeepFakes Attribution}
DeepFakes attribution technologies focus on extracting a unique fingerprint left by GMs within DeepFakes to determine its source model.
The pioneering work first introduced model fingerprints and designed the AttNet framework to trace the source model.
Building upon this foundation, subsequent research studies actively inserted transferable fingerprints into GMs, thereby enabling the decoupling of the fingerprint from the generated content ~\cite{DBLP:conf/iccv/YuSAF21,DBLP:conf/iclr/YuSCDF22}.
To enhance the capabilities of AMs in identifying unseen GMs, researchers explored DeepFakes attribution under an open-set setting ~\cite{DBLP:conf/iccv/GirishSRS21,DBLP:conf/cvpr/YangWTZCT23}.
Instead of attributing images to specific models, these studies aimed at architecture-level attribution, attributing images back to their source architectures ~\cite{DBLP:conf/icml/FrankESFKH20,DBLP:conf/aaai/YangHCL022,DBLP:conf/eccv/BuiYC22,DBLP:journals/pami/AsnaniYHL23}.
DCT revealed that architecture fingerprints cause severe artifacts in the frequency domain and proposed performing attribution within this domain ~\cite{DBLP:conf/icml/FrankESFKH20}.
DNA-Det identified the global consistency of architectural fingerprints and developed a patch-wise contrastive learning-based framework for attribution ~\cite{DBLP:conf/aaai/YangHCL022}. 
Meanwhile, some studies achieved this goal through a mixing representation strategy and reverse engineering, respectively~\cite{DBLP:conf/eccv/BuiYC22,DBLP:journals/pami/AsnaniYHL23}.
Recently, the rise of DMs spurred research into attributing images generated by these models.
For example, the image and its description are simultaneously utilized for DeepFakes detection and attribution~\cite{DBLP:conf/ccs/ShaLYZ23}.
Besides, reconstruction errors were used to infer the source model, as well-reconstructed images are likely generated from the inspected model ~\cite{DBLP:conf/icml/WangS0LMM24,DBLP:conf/icml/LaszkiewiczRLF24}.

\subsubsection{Anti Forensics}
DeepFakes forensics are critical for curbing misuse and establishing responsibility, driving many efforts devoted to exploring the vulnerabilities of existing forensic approaches, thus promoting more advanced forensic technologies, including DeepFakes detection and attribution.
Early anti-forensic work utilized adversarial examples like FGSM~\cite{DBLP:journals/corr/GoodfellowSS14}, PGD~\cite{DBLP:conf/iclr/MadryMSTV18} to add perturbations into DeepFakes for evading detectors ~\cite{DBLP:conf/cvpr/NeekharaDBC21,DBLP:conf/icip/LiaoLWKZLYSW21}.
DiffAttack subsequently used DMs to generate highly transferable perturbations ~\cite{10716799}.
Similarly, imperceptible semantic level perturbations were designed through latent space optimization~\cite{DBLP:conf/sp/MengWGJZ24}.
Some studies ~\cite{DBLP:conf/sp/WesselkampRAQ22,DBLP:journals/tdsc/LiuCZZZ23} evaded  detectors by reducing detectable artifacts in images rather than adding perturbations.
For instance, FakePolisher learned real representation from real images to construct a dictionary for reducing artifacts ~\cite{DBLP:conf/mm/HuangJWGMXLMLP20}.
StealthDiffusion ~\cite{DBLP:conf/mm/ZhouSCKSJ24} achieved this by optimizing both the latent space and the frequency domain, enabling images indistinguishable from real images.
% Similarly, ~\cite{DBLP:journals/tdsc/LiuCZZZ23} designed a framework where the generator is trained with adversarial training loss to optimize images' spatial, frequency domain, and fingerprint features simultaneously.
However, research on the vulnerability of AMs remains limited.
Specifically, the vulnerability faced with transformation-based methods like compression were explored ~\cite{DBLP:conf/iccv/Yu0F19,DBLP:conf/aaai/YangHCL022}.
Transferable adversarial samples were also attempted to attack AMs ~\cite{DBLP:conf/aaai/WuMWZLLL0024}.
TraceEvader, a universal attack method by inserting perturbations into high frequency information and blurring low frequency information of images, thus confusing traceable fingerprints to evade AMs~\cite{DBLP:conf/aaai/WuMWZLLL0024}.

Although existing works demonstrated considerable ASR, they fail to eliminate fingerprints within DeepFakes.
% which underscores the necessity for an attack method capable of bypassing current defensive mechanisms.
In this paper, we aim to propose a universal and black-box attack against DeepFakes AMs, more advanced, our method is more challenging to defend against.

\section{Preliminary Statement}
\subsection{Threat Model}
\subsubsection{Defender's goals and capabilities } 
The goal of the defender is to trace the source model of DeepFakes, determining which model instance or type of architecture the image is from.
The defender 1) trains attribution models with available clean DeepFakes and 2) attempts to alleviate the vulnerability of AMs through defensive strategies such as adversarial training. 
The defender may collect or create some adversarial DeepFakes for enhancing attribution.
% even if they have full knowledge of the attack model, including training data and model architecture.

\subsubsection{Adversary's goals and capabilities }
The goal of the attack is to generate DeepFakes and enhance untraceability to evade AMs even when faced with defensive measures, thereby avoiding responsibility.
The attacker 1) has full ownership of GMs and knowledge of training data, 2) has no access to or information about any AMs. 
3) aims to develop a universal attack that is agnostic to AMs, rendering untraceable DeepFakes without altering content perceptibly.

\subsection{Problem Formulation}
% 添加引用
Inspired by camera fingerprint studies \cite{DBLP:conf/www/QianHHMZY23}, an image $x$ generated by model $\mathcal{M}$ can be modeled as:
$$
        x =  \ x^0\ +\ x^0f_{\mathcal{M}}\ +\ \Theta 
$$
where $x^0$  represents the visual content of the image, $f_{\mathcal{M}}$ denotes the unique fingerprint left by model $\mathcal{M}$. In the following text, we use $f_{\mathcal{M}}$ to denote $x^0f_{\mathcal{M}}$ for brief.
$\Theta$ indicates other noise components.
Based on this, we provide definitions for attribution and attribution attack as follows:
\begin{definition}[Attribution]
    Let  $\mathcal{F}:\mathbb{R}^{\mathcal{CHW}} \rightarrow \mathcal{P(M)}$, where $\mathcal{P(M)}$ denotes the power set of all possible GMs, be an attribution model that identifies the source GMs $\mathcal{M}$ responsible for generating $x$. $\mathcal{F}$ aims to extract $f_{\mathcal{M}}$ from $x$ through
    $$\mathcal{F}(x) \mapsto f_{\mathcal{M}} \quad \text{s.t.} \quad  \|\mathcal{F}(x)-f_{\mathcal{M}}\| < \epsilon,$$
    where $\epsilon$ is a tolerance threshold for estimation errors.
\end{definition}

\begin{definition}[Attribution Attack]
    Let $\mathcal{T}:\mathbb{R}^{\mathcal{CHW}} \rightarrow \mathbb{R}^{\mathcal{CHW}}$ be an attack model that generates adversarial images to mislead $\mathcal{F}$ while preserving visual fidelity, that is
    $$
    || \mathcal{F}(\mathcal{T}(x))- \mathcal{F}(x) || \geq \epsilon \
    \text{and} \
    \mathcal{D}(\mathcal{T}(x),x) \leq \Delta,
    $$
    where $\mathcal{D}(\cdot, \cdot)$ is a metric bounding the visual distortion.
\end{definition}
This paper aims to design an attack method that is applicable to images generated by any GMs (universal) and capable of misleading all AMs without prior knowledge (black-box).

% Let $X=\{\langle x_1, y_1 \rangle,\langle x_2, y_2 \rangle,\langle x_3, y_3 \rangle \ldots,\langle x_n, y_n \rangle\}$ be a set of DeepFakes generated by a set of GMs $\{\mathcal{M}_{1},\mathcal{M}_{2},\mathcal{M}_{3},\ldots,\mathcal{M}_{m}\}$. Here, $y_i \in \{0,1,\ldots,m\}$ indicates the label of the source model for $x_i$.
% Given an image $x_i \in X$, 
% the goal of the attack method $\mathcal{T}$ is generating an adversarial images $x'_i=\mathcal{T}(x_i)$ with two constraints:
% $$
%     || \mathcal{F}(x'_i)- \mathcal{F}(x_i) || \geq \epsilon \
%     \text{and} \
%     \mathcal{D}(x',x) \leq \Delta
% $$
% where $\mathcal{D}(\cdot, \cdot)$ denotes the metric that quantifies the visual difference between the images.
% The former is an attack constraint to mislead the unknown AMs $\mathcal{F}(\cdot)$, and the latter is a visual constraint for ensuring the image's visual quality.

\subsection{Existing Attack Analysis}
In this section, we identify that existing attack methods are additive and fail to eliminate the fingerprint within DeepFakes, and thus are easily defended against. 

\begin{definition}[Additive Attack]
    Let $\mathcal{T}_{add}:\mathbb{R}^{\mathcal{CHW}} \rightarrow \mathbb{R}^{\mathcal{CHW}}$ be an additive attack model that generates adversarial images through inserting the perturbation into images:
    $$
    \mathcal{T}_{add}(x)=x+p
    $$
\end{definition}
In this context, a clean image $x$ generated is attacked as 
$$
    \mathcal{T}_{add}(x)= x^0 + f_{\mathcal{M}}\ + p + \Theta
$$
The success of additive attack methods lies in their ability to add a carefully crafted perturbation $p$ that obscures the fingerprint within DeepFakes, thereby misleading AMs.
It is important to note that this merely confuses the fingerprint, increasing the difficulty of extracting the fingerprint, however, the fingerprint itself remains intact within DeepFakes. 

We identify that existing attack methods are additive (Detailed analysis can be found in the Supplementary Material (SM) Sec 1), therefore, they are easily defended against as they fail to eliminate the fingerprint within DeepFakes.
Our defensive experiments and frequency analysis both reveal the limitations of existing attack methods: easily defended against and fail to eliminate fingerprints: 

\textit{1) Defensive Evaluations:}
We employ adversarial training to enhance the DNA-Det ~\cite{DBLP:conf/aaai/YangHCL022} and evaluate the effectiveness of additive attacks against defensive mechanisms.
 % 对抗训练细节放附录
% Specifically, during the adversarial training process, half of the original clean images in the training set are replaced with adversarial images.
Experimental results reveal that additive attacks can be easily and effectively countered by a defensive strategy.
As summarized in Table~\ref{tab:tab3}, the performance of all tested attacks experiences a significant decline following adversarial training.  For instance, TraceEvader's ASR drops from 98.28\% to 25.10\% against the defensive mechanism. 
% 补充下降幅度
% By incorporating adversarial images into the training dataset, defenders can significantly enhance the robustness of AMs. 
By incorporating adversarial training, defenders can significantly enhance the robustness of AMs to handle additive attacks.

% 指纹存在于高频，修改后可以观察到被修改了，和问题定义相对应，先说fre观察到指纹没有改变
\begin{figure}[tb]
\centering
  \includegraphics[width=\linewidth]{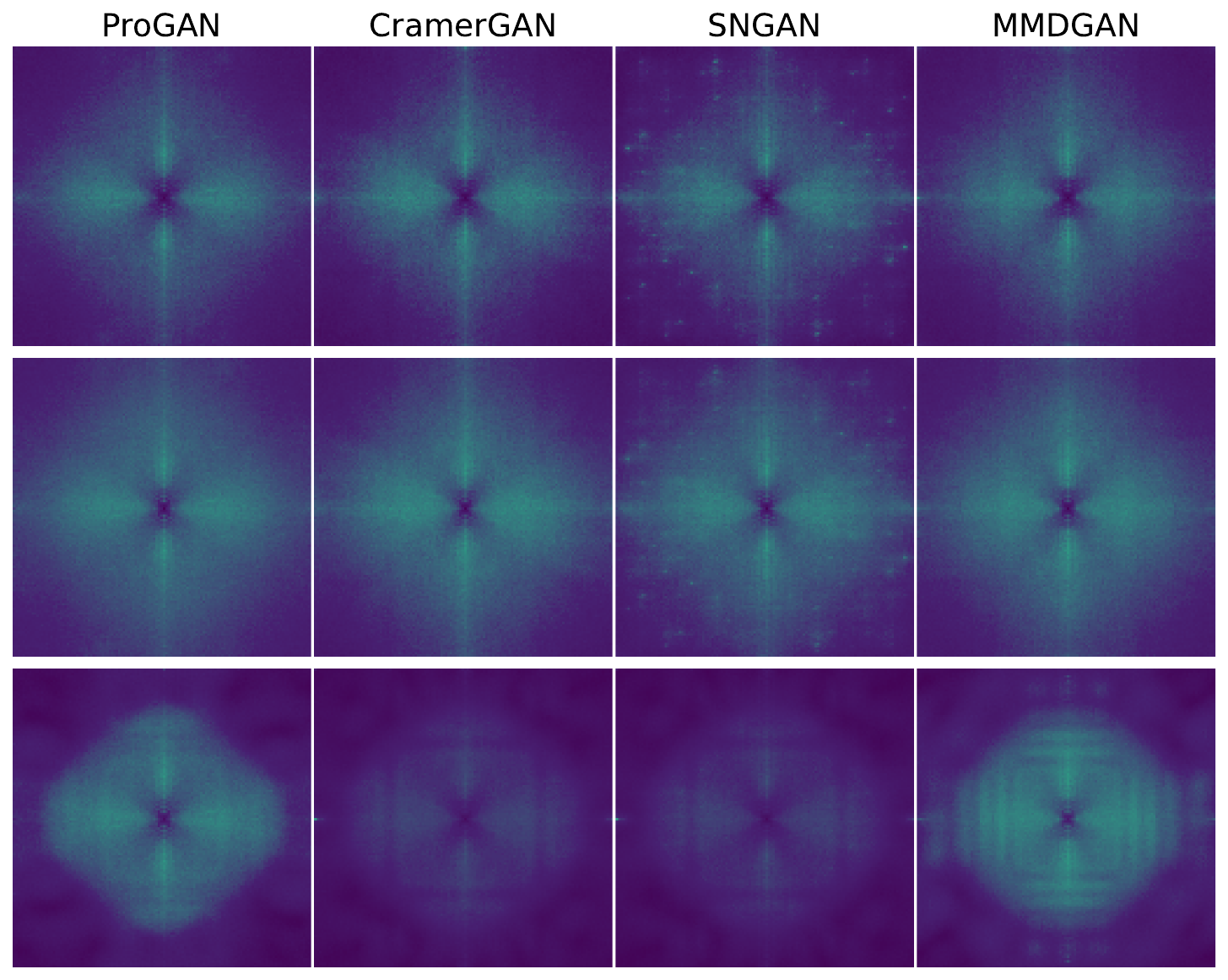}
  \caption{Spectral analysis of (top to bottom): original, additive attack images, and images processed with our method. 
  See SM Figures 6,7,8, and Sec 1.2 for more.
  }
  % \Description{}
  \label{fig:fig2}
\end{figure}

 \textit{2) Frequency Analysis:}
Model fingerprints exhibit pronounced characteristics in the frequency domain, manifesting as distinct patterns~\cite{DBLP:conf/icml/FrankESFKH20}. 
Our frequency analysis provides evidence of the persistence of these fingerprints within adversarial images.
As illustrated in Figure~\ref{fig:fig2}, images attacked by TraceEvader~\cite{DBLP:conf/aaai/WuMWZLLL0024} still exhibit high similarity in the frequency domain with their original counterparts, demonstrating that additive attacks fail to eliminate the fingerprint. More results are in SM Sec 1.2.

Therefore, merely confusing the traceable fingerprint is insufficient to truly evade AMs. 
Conversely, we argue that true non-tractability can only be achieved by eliminating the traceable fingerprint within the image.

% 具体的总结分析，放在附录
% The attack attempted to evade AMs can be summarized as follows:
% 1) Transformation-based attack: five types of transformation operations: noise, blurring, cropping, JPEG compression, relighting and random combination of them are considered to perturb images ~\cite{DBLP:conf/iccv/Yu0F19,DBLP:conf/aaai/YangHCL022}.
% 2) Transferable adversarial example: The BIM and MI-FGSM are leveraged to generate adversarial examples in a white-box setting against ~\cite{DBLP:conf/aaai/WuMWZLLL0024}, which are then transferred to other AMs in a black-box setting.
% 3) black-box and universal attack:  ~\cite{DBLP:conf/aaai/WuMWZLLL0024} adds a universal perturbation learned from DeepFakes. This perturbation is applied to the high-frequency components of images, while Gaussian blurring and mean shift are applied to the low-frequency components.

\section{Methodology}
In this section, we start by theoretically proposing our multiplicative attack and analyzing the difficulty of defending against the multiplicative attack.
And finally, we introduce the specific design of our multiplicative attack framework.

\subsection{Multiplicative Attack}

\begin{figure*}[tb]
\centering
  \includegraphics[width=\linewidth]{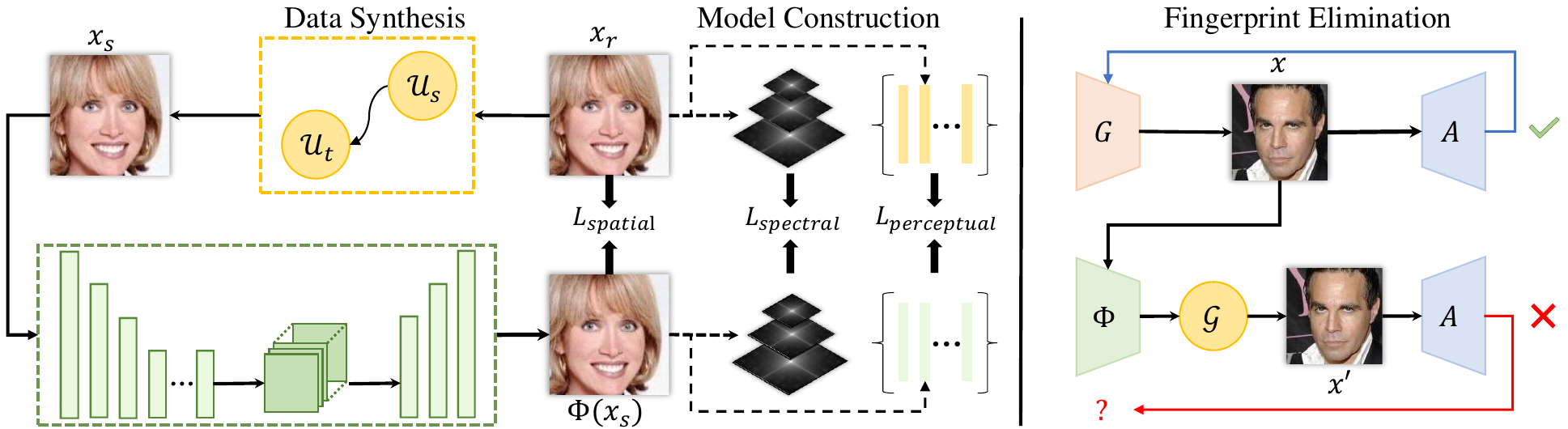}
  \caption{
The workflow of our proposed method. 
$\mathcal{U}_s$ and $\mathcal{U}_s$ represent the sample and transformation units.
$\Phi$ indicates our adversarial model, and $G$ and $A$ represent generative model and attribution model respectively.
$\mathcal{G}$ is smoothing operation.
  }
%   \Description{}
  \label{fig:fig1}
\end{figure*}

The principle of attribution methods is the extraction of the fingerprint $f_\mathcal{M}$ from DeepFakes. Therefore, we argue that true non-tractability can only be achieved by eliminating the model fingerprint within DeepFakes.
To achieve this, we propose a multiplicative attack that eliminates the traceable fingerprint through a multiplicative operation, rather than adding a perturbation $p$ to obscure the fingerprint:
\begin{definition}[Multiplicative Attack]
    Let $\mathcal{T}_{mul}:\mathbb{R}^{\mathcal{CHW}} \rightarrow \mathbb{R}^{\mathcal{CHW}}$ be a multiplicative attack model that generates adversarial images with an adversarial matrix $W$:
    $$
    \mathcal{T}_{mul}(x)=x\odot W 
    $$
\end{definition}
In this context, a clean image $x$ is attacked as:
$$
        \mathcal{T}_{mul}(x)=\ x^0 \odot W\ +\ f_{\mathcal{M}}\odot W\ +\ \Theta
$$
Here, $f_{\mathcal{M}}'=f_{\mathcal{M}} \odot W$ represents the altered fingerprint, making it distinct from the original $f_{\mathcal{M}}$.
Consequently, the source GMs can not be traced because $f_{\mathcal{M}}' \neq f_{\mathcal{M}}$.
By eliminating the original traceable fingerprint within DeepFakes, the multiplicative method ensures that no residual information related to the source GMs remains, thereby eliminating the possibility of exploitation for attribution, even those equipped with advanced defense mechanisms. 

% \textit{We provide an attack success lower bound for this attack, which yields Theorem \ref{the:the1}, see SM Sec.2 and Sec.3 for a detailed discussion.}

We prove that there always exists a multiplicative adversarial matrix $W$ capable of evading $\mathcal{F}$  while satisfying $||\Delta x||_2 =||x \odot (W-1)||_2 \leq \Delta$. Moreover, we establish a theoretical lower bound on the ASR,  which yields Theorem \ref{the:the1}, see SM Sec 2 and Sec 3 for a detailed discussion.
\begin{theorem}[Attack Success Lower Bound]
Under multiplicative attack $x' = x \odot W$ with $\|x\|_\infty \leq 1$ and $\|x \odot (W-1)\|_2 \leq \Delta$, if $\mathcal{F}$ is differentiable almost everywhere and correctly classifies $\mathcal{D}$-almost every $x$, then for sufficiently small $\Delta$:
$$
\text{ASR}(\Delta) \geq \Pr_{x \sim \mathcal{D}}\left[\gamma_{\text{attack}}(x) \geq \frac{2\Delta_{\text{conf}}(x)}{\Delta}\right]
$$
where $\Delta_{\text{conf}}(x)$ is the confidence gap and $\gamma_{\text{attack}}(x)$ measures vulnerability to multiplicative perturbations. When this condition holds, attack succeeds with certainty.
\label{the:the1}
\end{theorem}

\subsection{Defense Difficulty Analysis}
One potential defensive strategy is to invert the adversarial image $x'$ to obtain the original image $x=x' \odot W^{-1}$.
However, solving $W$ requires a large number of pairs of a clean image $x$ and a corresponding adversarial image $x'$.
Besides, inverting the matrix $W$ would require a computational complexity of $O(n^3)$, which is computationally prohibitive and unaffordable.
Even if all the elements within matrix $W$ are binary (0 or 1), the probability of correctly guessing $W$ is $\frac{1}{2^{d}}, W \in \mathbb{R}^{d}$. This probability is astronomically low, making brute-force guessing infeasible.
Given the  impractical data requirements and prohibitive computational complexity,  it is fundamentally infeasible for a defender to obtain $W^{-1}$. 

Another potential strategy is using a neural network to approximate the matrix $W^{-1}$, thereby attempting to enhance AMs.
However, this also requires the defender to collect a large number of image pairs, which is impractical.
Moreover, using a network to invert images imprints its fingerprints onto the recovered content, further degrading defense efficacy.
Our experiments show that even with simultaneous access to both clean and adversarial images, defenders still cannot reliably attribute the source of DeepFakes.
The intrinsic disruption of the model fingerprint within the images renders such defensive strategies ineffective.

\subsection{Multiplicative Adversarial Attack Framework}
% 需要和理论承接
The primary objective of this framework is to construct an adversarial model that serves as the adversarial matrix to eliminate fingerprints within DeepFakes, thus evading AMs, even under enhanced defensive measures.
The proposed framework is universally applicable across diverse GMs without requiring customization and remains effective against various AMs under black-box scenarios.

\subsubsection{Overview} 
Our framework is explicitly designed for end-to-end adversarial model construction without relying on any generative data, GMs or access to specific AMs. 
% This model acts as the adversarial matrix $W$ to eliminate the traceable fingerprint within the image through a multiplicative operation, enabling the fundamental non-traceability of DeepFakes.
As shown in Figure~\ref{fig:fig1}, our framework comprises a data synthesis module, a model construction module and a fingerprint elimination module. 
The data synthesis module is designed for the creation of synthetic data via inserting artificial fingerprints into real data, mimicking the characteristics of data generated by GMs.
In the model construction module, the attack model learns to eliminate artificial fingerprints within synthetic data while preserving the visual perception of the images.
After training, this model functions as the matrix $W$ to eliminate the traceable fingerprint left by the corresponding GMs, thereby creating untraceable Deepfakes.

\subsubsection{Insight}
The following motivation drives our framework to mimic the fingerprints of GMs:
\textit{1) Sampling Operations in GMs}: GMs primarily rely on down- and up-sampling operations to generate images, such as nearest neighbor up-sampling, which introduces grid-like patterns in both the spatial and frequency domains ~\cite{Odena_Dumoulin_Olah_2016,DBLP:conf/icml/FrankESFKH20}. These patterns are inherent artifacts of the sampling processes used by GMs.
\textit{2) Image Transformation Similarities}: Certain image transformation operations exhibit characteristics similar to those found in GMs. For example, blurring and adding noise using kernels closely resemble the convolution computations performed~\cite{DBLP:conf/aaai/YangHCL022}.
These transformations share spatial properties with the operations conducted within GMs.
Therefore, the traces left by sampling and transformation operations exhibit properties that are analogous to the fingerprints of GMs.

\subsubsection{Data Synthesis}
To effectively mimic the fingerprints of GMs, we design sampling and transformation units within the data synthesis module, as illustrated in Figure~\ref{fig:fig1}.   
The sampling unit $\mathcal{U}_{s}(\cdot)$ employs three sampling techniques: nearest-neighbor, bilinear, and bi-cubic interpolation. 
Given a real image $x_{r} \in \mathbb{R}^{\mathcal{CHW}}$, it is first down-sampled to $x_{down} \in \mathbb{R}^{\mathcal{C'H'W'}}$, where $\mathcal{C'=C}/2, \mathcal{H'=H}/2$, and $\mathcal{W'=W}/2$.
And then the image is up-sampled back to its original dimensions, resulting in $x_{up} \in \mathbb{R}^{\mathcal{CHW}}$.
Specifically, the sampling unit stochastically applies down/up-sampling with probability $p_1$: $x_{up}=\mathcal{U}_{s}(x_{r},s_{down},s_{up},p_{1})$, where  $s_{down},s_{up}$ are selected randomly to introduce diverse spatial artifacts and enhance robustness. 

The transformation unit $\mathcal{U}_{t}(\cdot)$ incorporates a series of image transformation techniques designed to introduce diverse and realistic variations, and the details are as follows:
% Gaussian noise addition: Adds Gaussian noise sampled from $\mathcal{N}(0,\sigma^2)$, where the variance $\sigma^2$ is randomly selected from the interval [5.0, 20.0].
% Blurring: Applies Gaussian filtering with a kernel size randomly chosen from\{1,3,5\}.
% Cropping: Randomly crops the image with an offset between 5\%-20\% of the image side lengths.
% JPEG compression: Performs JPEG compression using a quality factor randomly sampled from [10,75].
% Random relighting: Adjusts brightness, contrast, and saturation by sampling random factors within the range [0.5, 1.5].
(a) Gaussian noise sampled from $\mathcal{N}(0,\sigma^2)$, where $\sigma^2$ is randomly selected from  [5.0, 20.0];
(b) Gaussian filtering with a kernel size randomly chosen from\{1,3,5\};
(c) Randomly crops with an offset between 5-20\% of the image lengths;
(d) JPEG compression using a quality factor randomly sampled from [10,75];
(e) Relighting (adjusts brightness, contrast, and saturation) with random factors from [0.5, 1.5];
(f) Combination processes in the following order: relight, cropping, blurring, compression and adding noise.
For each image, one of the above operation $t$ is randomly selected and applied with probability $p_2$: $x_s= \mathcal{U}_{t}(x_{up},t,p_2)$.

By mimicking these properties of sampling and transformation operations, we can effectively simulate the characteristics of GM-generated data without access to specific GMs.

\subsubsection{Model Construction}
The adversarial model $\Phi$ is trained end-to-end on real/synthetic image pairs ($x_r,x_s$) to eliminate artificial fingerprints within synthetic images while preserving perceptual fidelity.
% 介绍模型结构，在附录进行细说，gbms
$\Phi$ comprises an encoder-decoder architecture. The encoder includes 3 convolutional layers and 5 residual layers to extract feature maps from the input images. 
The decoder utilizes 2 up-sampling layers and a convolutional layer to reconstruct images from feature maps. 
The details are presented in SM Sec 4 and Table 3.
% Specifically, the decoder employs three down-sampling techniques—nearest-neighbor, bi-linear, and bi-cubic interpolation—with the specific type of sampling randomly selected during training.

To ensure visual and semantic similarity between  $\Phi(x_s)$ and $x_r$, we employ a perceptual loss function with a pretrained 16-layer VGG network as the backbone to extract high-level features. 
The perceptual loss is defined as:
$$
\mathcal{L}_{perceptual}=\sum_{i\in F}w_{i}||f_{\Phi(x_s)}^{i}-f_{x_r}^{i}||_{2},
$$
where $f_{x}^i$ denotes the feature extracted from the selected layer $F=\{f^{i_1},f^{i_2} \dots f^{i_j}\}$ of the perceptual network for image $x$, and $w_i=\frac{1}{|F|}$ represents the weight.
To mitigate low-level artifacts within $x_s$, we define a spatial loss to remove these artifacts further to preserve the image's detail:
$$
    \mathcal{L}_{spatial}= 
    ||\Phi(x_{s}) - x_{r}||_{2}.
$$
Given that synthesis operations and GMs introduce artifacts in both spatial and frequency domains, we design a multi-scale spectral loss for frequency domain optimization.
Specifically, we apply the Fourier Transform to obtain the frequency representation of a scaled image $x_{s}$ with scale factor $s\in S=\{1, 0.5, 0.25\}$, calculating the magnitude of the frequency components, and applying logarithmic scaling to stabilize numerical computations. 
And a small constant $\varepsilon$ is added to prevent taking the logarithm of zero values.
That is 
$$
    \mathcal{L}(x,s_i)=log(|\textit{fft}(x^{s_i})|+\varepsilon).
$$
The final spectral loss is formulated as
$$
    \mathcal{L}_{spectral}= \sum_{s_{i}\in S} w_i
    ||\mathcal{L}(\Phi(x_{s}),s_i)- \mathcal{L}(x_{r},s_i)||_{1},
$$
where $w_i\in \{0.5, 0.3, 0.2\}$ represents the weight.
The total training loss function is formulated as:
$$
\mathcal{L}_{total}=\beta_1\mathcal{L}_{perceptual}+\beta_2 \mathcal{L}_{spatial} + \beta_3 \mathcal{L}_{spectral},
$$
where $\beta_1,\beta_2,\beta_3$ are set as $\{0.5, 0.1, 0.4\}$.

Through jointly optimizing in the perceptual, spatial, and frequency domains, the trained attack model $\Phi$ can effectively eliminate artificial traces introduced by sampling and transformation operations while preserving visual similarity with the original image. 

\subsubsection{Fingerprint Elimination} After training, given a clean and traceable generative image $x$, the attack model $\Phi$ eliminates the fingerprints left by the corresponding source GMs, thereby achieving an untraceable adversarial image for evading DeepFakes attribution.
Furthermore, to eliminate imperfection or distortion artifacts and enhance the evasion capability, we implement a hybrid image smoothing operation combining Gaussian Blur and Mean Shift filtering (GBMS) $\mathcal{G}(\cdot)$.
The adversarial image is generated as:
$$
    x'=\mathcal{T}_{mul}(x|\Phi)=\mathcal{G}(\Phi(x))
$$

\section{Experiments}

In this section, we conduct comprehensive experiments to evaluate the attack performance to evade AMs and against the defensive mechanisms of our attack.
Besides, we carry out a quantitative analysis on both image and frequency to demonstrate that our attack method is multiplicative and can effectively eliminate the fingerprints within images.

\subsection{Experiments Setup}
We conduct experiments spanning 6 GANs, 2 DMs, and 3 datasets against 6 advanced AMs, and implement 8 attack methods, including 4 transferable attacks (PGD, BIM, MIFGSM, DiffAttack~\cite{10716799}), 3 black-box methods (transformation, FakePolisher~\cite{DBLP:conf/mm/HuangJWGMXLMLP20}, TraceEvader~\cite{DBLP:conf/aaai/WuMWZLLL0024}), and 1 regeneration attack~\cite {DBLP:conf/nips/ZhaoZSVGKVWL24}.
Details in SM Sec 5.1, Table 4.

\begin{table*}[tb]
\begin{center}
% \scalebox{0.85}{
\begin{tabular}{lccccccccc}
\toprule
               & \multicolumn{1}{|c}{DNA-Det} & AttNet & DCT   & Reverse & POSE &LTracer & \multicolumn{1}{|c|}{Average} &SSIM & LPIPS\\ 
\midrule
PGD            & \multicolumn{1}{|c}{-}       & \underline{99.98}  & 66.16 & 59.77   & 6.75  & - & \multicolumn{1}{|c|}{58.17} &0.912  &0.401\\
BIM            & \multicolumn{1}{|c}{-}        & 0.75   & 47.25 & 43.78   & 3.40  & - & \multicolumn{1}{|c|}{23.79} &0.910  &0.401\\
MIFGSM         & \multicolumn{1}{|c}{-}        & 8.90    & 80.25 & 67.58   & 1.30 & -  & \multicolumn{1}{|c|}{39.51} &0.911  &0.401\\
DiffAttack    & \multicolumn{1}{|c}{-}        & \textbf{100.00}    & 66.75 & 25.40    & 56.8  & - & \multicolumn{1}{|c|}{62.24} &0.962  &0.095\\
% \hline
Transformation & \multicolumn{1}{|c}{43.70}   & 99.96  & 61.21 & 54.89   & 47.01 & \underline{98.79} & \multicolumn{1}{|c|}{67.60} &0.941  &0.151\\
FakePolisher   & \multicolumn{1}{|c}{75.00}    & 40.95  & 53.70  & \textbf{90.33}   & \textbf{95.85}  & - & \multicolumn{1}{|c|}{71.17} &\underline{0.994}  &\underline{0.067}\\
Regeneration   & \multicolumn{1}{|c}{93.30}    & 99.95  & 84.46  & 81.05   & 39.71  & 0.00 & \multicolumn{1}{|c|}{78.60} &0.912  &0.210 \\
TraceEvader    & \multicolumn{1}{|c}{\underline{98.28}}   & 65.45  & \underline{92.40}  & 88.80    & 86.81  & 90.89 & \multicolumn{1}{|c|}{\underline{87.11}} &\textbf{0.995}   &\textbf{0.038}\\
    Ours      & \multicolumn{1}{|c}{\textbf{98.56}}       & \textbf{100.00}    & \textbf{100.00}   & \underline{89.54}       & \underline{95.52}  & \textbf{98.89}   & \multicolumn{1}{|c|}{\textbf{97.08}} &0.963  &0.093\\

\midrule
Ours (w/o $\mathcal{U}_s$)      & \multicolumn{1}{|c}{98.10}       &100.00    &100.00    &84.75  & 93.76  &-   & \multicolumn{1}{|c|}{95.32} &0.970  &0.076\\
Ours (w/o $\mathcal{U}_t$)      & \multicolumn{1}{|c}{91.69}       &100.00    &100.00    &81.48  &95.83   &-   & \multicolumn{1}{|c|}{93.80} &0.968  &0.083\\
Ours (w/o GBMS)      & \multicolumn{1}{|c}{86.15}       &100.00    &100.00    &67.83  &95.14   &-   & \multicolumn{1}{|c|}{89.82} &0.965  &0.102\\
Ours (w/o $\mathcal{L}_{spatial}$)      & \multicolumn{1}{|c}{86.34}       &100.00    &100.00    & 87.80 &96.89   &-   & \multicolumn{1}{|c|}{89.82} &0.964  &0.102\\
Ours (w/o $\mathcal{L}_{spectral}$)      & \multicolumn{1}{|c}{92.45}       &100.00    &61.16    &50.04  &97.92   &-   & \multicolumn{1}{|c|}{80.31} &0.950  &0.089\\
\bottomrule
\end{tabular} 
% }
\caption{Comparison of ASR (\%) across various AMs. The best and second-best results are marked in \textbf{bold} and \underline{underlined} (except for the ablation part). The symbol `-' within the table indicates not applicable or limited by computational overhead.}
\label{tab:tab2}
\end{center}
\end{table*}

\begin{table*}[]
\begin{tabular}{c|cccccccc}
\toprule
          & PGD   & BIM   & MIFGSM & Transformation & FakePolisher & Regeneration & TraceEvader & Ours   \\ \midrule
Black-box & 81.34 & 77.18 & 82.16  & 76.13          & 91.43        & \textbf{100.00}       & 72.39       &  \\
White-box & 0.02  & 0.00  & 0.00   & 28.53          & 9.63         & \textbf{100.00}       & 25.10       & \textbf{100.00} \\ \bottomrule
\end{tabular}
\caption{Evaluation results against defensive mechanisms (adversarial training).  The best results are marked in \textbf{bold}.
    }
\label{tab:tab3}
\end{table*}

\subsection{Effectiveness Against Attribution Methods}
In this section, we evaluate the effectiveness of our proposed method in evading DeepFakes attribution technologies and demonstrate the broad applicability and universality of our approach across diverse GMs, including GANs and DMs.

% Transferable attacks are applied exclusively to DeepFakes generated by ProGAN to evade POSE.
Our method demonstrates the superior effectiveness of evading AMs in a black-box setting and broad universality across different types of GMs.
As shown in Table~\ref{tab:tab2}, it achieves the highest ASR in attacking four AMs and the second-best ASR in two AMs, significantly outperforming existing SOTA techniques. Notably, our method attains an average ASR of 97.08\%, markedly higher than the 87.11\% achieved by TraceEvader.
% regeneration
While the regeneration attack achieves considerable ASR in partial AMs,
its performance remains suboptimal compared to our approach. Particularly, 
its effectiveness degrades to 39.71\% and 0.0\% against POSE and LTracer, respectively, highlighting the necessity of fingerprint elimination for attack.
% 消除扩散模型指纹
Besides, our method achieves an ASR of 98.89\% against LTracer, validating the efficacy in eliminating DMs' fingerprints.

% The results also indicate that our method can effectively eliminate fingerprints of both GANs and DMs, underscoring the universality and versatility of our approach in diverse GMs contexts.

% 困惑矩阵
We provide visual confusion matrices to quantitatively characterize the adversarial attack performance in SM Figure 9 and Sec 5.2. Empirical results indicate that our method effectively induces AMs to misclassify DeepFakes as real images, thereby validating the perceptual indistinguishability between adversarial images and real images. 

% In summary, these evaluation not only highlight the effectiveness of our method in a black-box setting but also demonstrate its wide-ranging applicability across different types of GMs, making it a powerful tool for challenging DeepFakes attribution technologies.

\subsection{Effectiveness Against Defensive Mechanisms}

% \begin{table}[tb]
% \begin{center}
% \begin{tabular}{cccl}
% \toprule
% AMs       & ASR & AMs                   & ASR \\ 
% \toprule
% $\mathcal{F}_{PGD}$     & 100.00   & $\mathcal{F}_{FakePolisher}$   & 100.00  \\
% $\mathcal{F}_{BIM}$    & 100.00    & $\mathcal{F}_{TraceEvader}$    & 99.91    \\ 
% $\mathcal{F}_{MIFGSM}$ & 100.00   & $\mathcal{F}_{Our}$    & 87.89    \\ 
% \bottomrule
% \end{tabular}
% \caption{Attack success rates against defensive mechanisms.
%     $\mathcal{F}_{PGD}$ is the DNA-Det enhanced with adversarial images created by PGD.
%     }
% \label{tab:tab3}
% \end{center}%\vspace{-5mm}
% \end{table}

In this section, we showcase the superior effectiveness of the proposed method against defensive mechanisms, even when the defender has full knowledge of our adversarial model. 
% This highlights the challenges in defending our approach.

To better assess our adversarial model against the adaptive defense, we identify two distinct scenarios:
\textit{1) Black-Box Scenario}: the defender is unaware of the existence of our attack and can not directly utilize our attack model to generate adversarial images for improving the robustness of AMs.
However, the defender can attempt to enhance AMs by creating adversarial images through all other known attacks. 
\textit{2) White-Box Scenario}: the defender is fully aware of our method and  can directly utilize adversarial images created by the proposed method to enhance AMs.

\textit{1) Against Adversarial Training:}
Experimental results demonstrate that neither white-box nor black-box defenses can effectively resist our attacks through adversarial training. As shown in Table \ref{tab:tab3}, our attack achieves over 72.39\% ASR in black-box scenarios. Even when defenders directly use our adversarial images to enhance DNA-Det, no mitigation effect is observed, the ASR even reaches 100\%.

\textit{2) Against Approximate Inversion:}
We demonstrate that training deep neural networks to invert our adversarial image and recover the original image is infeasible. Two models that are based on the DnCNN and the AutoEncoder architectures are trained under image pair supervision and evaluated on new adversarial samples. The inverted images maintain an ASR of 97.68\% and 99.97\%, respectively, validating the defensive resistance of our method. Details in SM Sec 5.2.

\subsection{Effectiveness Preserving Image Fidelity}
To evaluate the imperceptibility of our attack, we employ the SSIM and LPIPS metrics. As shown in Table \ref{tab:tab2}, our method achieves comparable performance to TraceEvader in maintaining image quality, yielding 0.963 SSIM and 0.093 LPIPS values. We provide a visual comparison between original and adversarial images in SM Sec 5.2 and Figure 11, demonstrating the ability of our approach in preserving quality.

\subsection{Ablation Study}
Ablation studies validate the importance of components ($\mathcal{U}_s, \mathcal{U}_t,$ GBMS) and function design ($\mathcal{L}_{spatial}, \mathcal{L}_{spectral}$) in our framework. As shown in Table \ref{tab:tab2}, while removing individual components does not degrade image quality, it significantly reduces ASR. For instance, eliminating GBMS causes ASR to drop from 97.08\% to 89.82\%. Notably, without $\mathcal{L}_{spectral}$, our method achieves only 50.04\% ASR against Reverse AMs. These results conclusively demonstrate the necessity of all designed components and loss functions.

\subsection{Quantitative Analysis}

We conduct quantitative analysis on the frequency and image domains,  demonstrating that our attack can effectively eliminate the traceable fingerprint within DeepFakes, and it is truly multiplicative rather than additive.

\subsubsection{Frequency Domain Analysis}
% We compute the two-dimensional Fast Fourier Transform of images, shift the zero frequency component to the center of the spectrum and calculate the logarithm of the absolute values of the shifted spectrum to obtain the magnitude spectrum.
Analysis results indicate that our attack can effectively eliminate the fingerprint within DeepFakes.
As shown in Figure~\ref{fig:fig2}, images attacked by our method exhibit significant differences from the original ones in the frequency domain. This visual evidence indicates that our attack successfully eliminates the source model's fingerprint from DeepFakes.
See Figures 6,7 for more.
% This result indicates that additive attack can not effectively eliminate the fingerprint of GMs , they only increase the difficulty of fingerprint extraction through confusing, while this is achieved by our multiplicative attack.

\subsubsection{Image Domain Analysis} 
\begin{figure}[tb]
  \includegraphics[width=\linewidth]{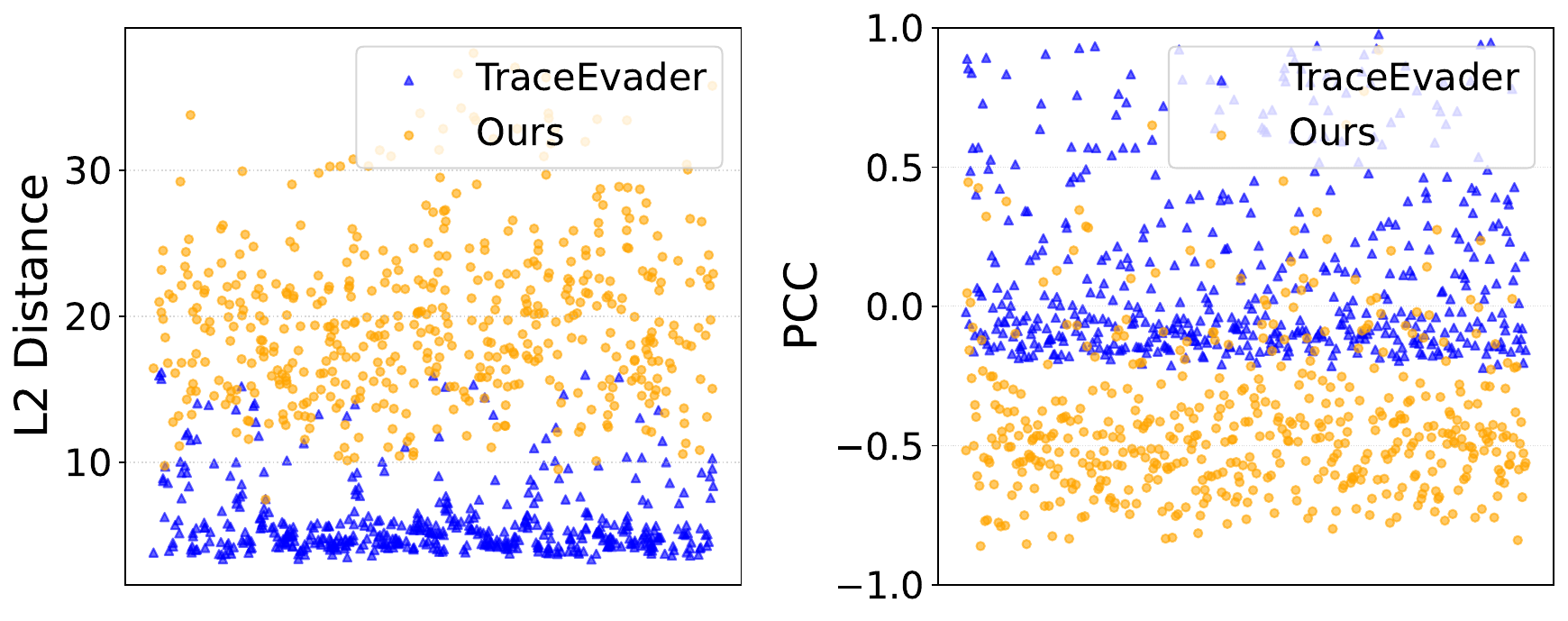}
  \caption{
Distribution of L2 distances between $\Delta$ (left) and Pearson Correlation Coefficient (PCC) between $\Delta$ and original images (right). See SM Figure 4, 5 and Sec1.2 for more.
  }
%   \Description{}
  \label{fig:fig3}
\end{figure}

Distance and correlation analysis both demonstrate that the proposed attack method exhibits multiplicative characteristics rather than additive:

1) \textit{L2 distance among Residual Components:}
The residual component $\Delta = \mathcal{T}(x)-x$ verifies attack nature:
fixed/stable $\Delta$ indicates additive attacks, while variable $\Delta$ signifies multiplicative behavior.
As visualized in Figure~\ref{fig:fig3}, our attack exhibits significantly higher variance than TraceEvader's stable residuals, confirming its multiplicative characteristics.

2) \textit{Correlation between $\Delta$ and Original Images:}
The residual components caused by multiplicative attack $\Delta=x\odot W -x =x\odot(W-1)$ exhibit higher correlation with original images than those caused by additive attack $\Delta=p$.
% We use Pearson Correlation Coefficient (PCC) to measure the correlation, the absolute value of PCC closer to 1 indicates stronger the correlation.
As shown in Figure~\ref{fig:fig3}, the $\Delta$ caused by our method exhibits high correlation with original images, with the absolute value of PCC mainly ranging from 0.5 to 1.0, while those caused by TraceEvader mainly range from 0.0 to 0.25.

\section{Conclusion and Limitation}
In this paper, we analyze the principles underlying existing attack methods, revealing a critical structural limitation: current attacks are additive and thus easily defended against because they fail to eliminate GMs' fingerprints.
We propose that the true non-traceability in DeepFakes can be achieved by the multiplicative attack, and provide theoretical evidence that this attack is irreversible and poses a significant challenge for defensive strategies.
Building on this foundation, we design a universal attack method where the adversarial model learns to eliminate GMs' fingerprints using only real data. 
This approach ensures broad applicability across diverse GMs, without requiring any knowledge of AMs.
Experimental results demonstrate that our method achieves an ASR of 97.08\% across 6 advanced DeepFakes attribution technologies and 9 generative models, surpassing SOTA methods. 
Additionally, it exhibits significant effectiveness with an ASR exceeding 72.39\% against defensive mechanisms.
Our research highlights the significant challenges introduced by multiplicative attacks and underscores the necessity for developing more advanced attribution mechanisms.

While our method achieves significant success in eliminating model fingerprints, it necessitates more modifications compared to additive attacks. Although these changes do not impact visual perception, we remain committed to optimizing our approach in future work and exploring more advanced and efficient attack and defense strategies.

\bibliography{aaai2026}

\end{document}